\documentstyle[12pt]{article}
\begin{document}

\begin{titlepage}
\noindent
\begin{flushright}
PRA -- HEP 97/2\\
January 1997\\
hep--ph/9701397
\end{flushright}

\vfill

\begin{center}
\noindent
{\huge\bf{Dispersive derivation of the trace anomaly}}

\vspace{1cm}

\noindent
{\large
   J. Ho\v{r}ej\v{s}\'{\i}\footnote{e-mail:Jiri.Horejsi@mff.cuni.cz}
   and M. Schnabl\footnote{e-mail:schnabl@hp01.troja.mff.cuni.cz}}
   \\
{\it Nuclear Centre, Faculty of Mathematics and Physics, Charles
University}\\
{\it V Hole\v{s}ovi\v{c}k\'ach 2, CZ-180 00 Prague 8, Czech Republic }

\end{center}

\vfill

\begin{center}
{\bf Abstract}
\end{center}

We present a simple derivation of the one-loop trace anomaly in spinor QED through dispersion
relations, avoiding completely any ultraviolet regularization. The anomaly can be expressed as a
convergent sum rule for the imaginary part of a relevant formfactor. In the massless limit, the
imaginary part produces a delta-function singularity at zero external momentum squared. Such a
treatment reveals an "infrared face" of the trace anomaly, in striking similarity with the well-known
case of the axial anomaly.

\noindent

\vfill

\end{titlepage}

\newcommand{\im}{\mathop{\rm Im}\nolimits}
\let\eps = \varepsilon

\section{Introduction}

Quantum anomalies constitute one of the fundamental chapters in modern field theory and play a
remarkably ubiquitous role in particle physics \cite{B}. A thorough understanding of their various
aspects therefore represents an important goal of the theory. In the archetypal example - the axial
anomaly \cite{ABJ} - one observes a peculiar feature: The relevant quantity can be calculated in several
clearly distinct ways, which in turn reveal diverse faces of the anomaly. To put it in more explicit
terms, let us recall that the most familiar calculational methods refer to the ultraviolet properties of
the basic triangle graph (exemplified by the Pauli-Villars or dimensional regularization). On the
other hand, the axial anomaly can be computed in a radically different way, namely by using
(convergent) dispersion relations and some remarkable infrared properties of the relevant
imaginary part \cite{DZ,FHVT}. Thus, one may avoid completely a regularization of the ultraviolet
divergences associated with the triangle graph. This "infrared face" of the axial anomaly is
relatively rarely mentioned in the literature, in comparison with the ultraviolet approaches
discussed amply in any modern field theory textbook (see, however,
\cite{B,H} ). Nevertheless, it represents a conceptually independent,
complementary view of the anomaly phenomenon.

  In this context, another potentially interesting example would be the well-known trace anomaly
\cite{CJ,ChE,ACD}, related to the broken dilatation (scale) invariance. The trace anomaly (which manifests itself
as an additional quantum contribution to the trace of the energy-momentum tensor) has been first
calculated by using an explicit ultraviolet regularization of the relevant correlation functions, and
interpreted in terms of the corresponding high-momentum asymptotic behaviour
\cite{CJ,ChE}. Subsequently
it has been related to the Gell-Mann-Low beta function associated with coupling constant
renormalization \cite{ACD}, and this "ultraviolet face" of the trace anomaly has become a common
wisdom in modern particle theory. It would be interesting to know whether the trace anomaly
could also be recovered through dispersion relations (and related to infrared properties of the
imaginary part of an appropriate correlation function), in analogy with what has been done before
in the case of the axial anomaly \cite{DZ,FHVT}. To our knowledge, there
has been no detailed discussion of this point in the available
literature, although the trace anomaly already has a rather
long history. The purpose of the present paper is to examine this problem explicitly, employing the
familiar framework of canonical Ward identities (zero-energy theorems) for broken scale
invariance. To fix a reference point for the subsequent discussion and for a later comparison, in the
next section we summarize briefly some basic results concerning the dispersive approach to the
axial anomaly (cf. \cite{B,DZ,FHVT,H} ). The dispersive derivation of the trace anomaly within spinor QED
is given in Section 3, which in fact represents  the central part of the paper. Some other aspects of
this approach are discussed in Section 4 and the main results are summarized briefly in Section 5.

\section{Paradigm of the axial anomaly}

Consider the familiar $VVA$ triangle fermion loop and denote the
corresponding (properly normalized) amplitude as
$T_{\alpha\mu\nu}(k,p)$, with $k, p$ being the
external momenta outgoing from vector vertices (labelled by $\mu, \nu$);
for simplicity, we take $k^2 = p^2 = 0.$ One may introduce formfactors via
an appropriate tensor decomposition as
\begin{equation}
T_{\alpha\mu\nu}(k,p) = F_1(q^2) q_\alpha
\eps_{\mu\nu\rho\sigma} k^\rho p^\sigma
+F_2(q^2) (p_\nu \eps_{\alpha\mu\rho\sigma} - k_\mu
\eps_{\alpha\nu\rho\sigma}) k^\rho p^\sigma
\end{equation}
Note that in the last expression we have already imposed the vector Ward identities (i.e. vector
current conservation). The associated $VVP$ amplitude can be written as
\begin{equation}
T_{\mu\nu}(k,p) = G(q^2) \eps_{\mu\nu\rho\sigma} k^\rho p^\sigma
\end{equation}
and the (anomalous) axial  Ward identity then reads
\begin{equation}
q^2 F_1 = 2m G + \frac{1}{2\pi^2}
\end{equation}
where $m$ denotes the fermion mass; the constant $1/2\pi^2$ is
the celebrated axial anomaly \cite{ABJ} (we have suppressed here any
possible coupling constants). If one wants to reproduce the anomaly through
the corresponding imaginary parts (following [3,4] ), one may define the formfactors in (1) and (2)
by means of dispersion relations (which turn out to converge without subtractions). The imaginary
parts are well-defined finite quantities and must therefore satisfy the canonical (i.e. non-
anomalous) Ward identities, i.e.
\begin{equation}
t\im F_1(t;m^2) = 2m \im G(t;m^2)
\end{equation}
where $t$ stands for the kinematical variable $q^2$ (we have marked
explicitly also the parametric dependence on the fermion mass). Using
unsubtracted dispersion relations for the $F_1$ and $G$ and
employing the identity (4), one gets readily
\begin{equation}
q^2 F_1 = 2m G - \frac{1}{\pi}
\int_{4m^2}^{\infty}\im F_1(t;m^2) dt
\end{equation}
An explicit calculation gives
\begin{equation}
\im F_1(t;m^2) = -\frac{1}{\pi}\frac{m^2}{t^2}
\ln\frac{1+\sqrt{1-\frac{4m^2}{t}}}{1-\sqrt{1-\frac{4m^2}{t}}}
\end{equation}
for $t>4m^2$, and taking the integral one obtains
\begin{equation}
\int_{4m^2}^{\infty}\im F_1(t;m^2) dt = -\frac{1}{2\pi}
\end{equation}
Using (5) and (7), the value of the axial anomaly in (3) is recovered. In such a derivation one thus
obviously avoids any ultraviolet regularization and the anomaly emerges as a sum rule for the
imaginary part of a relevant formfactor (which is simply related to the classical symmetry-breaking
term - cf. (4) ). In the massless limit, the $\im F_1(t;m^2)$ is seen to vanish pointwise while the integral
remains constant, independent of $m$. It means that, in fact
\begin{equation}
\lim_{m\to 0} \im F_1(q^2;m^2) = -\frac{1}{2\pi} \delta(q^2)
\end{equation}
In the approach outlined above, the fermion mass $m$ serves as an infrared cut-off and the anomaly,
being the net effect persisting in the massless limit, is a consequence of a delta-function singularity
of the $\im F_1$ at $q^2 = 0$ (in the full formfactor $F_1$ it is
manifested as an "anomaly pole" at $q^2 = 0$).
These facts, in a nutshell, describe the infrared (or low-energy) face of the axial anomaly.

\section{Trace anomaly in spinor QED through \\
         dispersion relations}

Let us now examine the trace anomaly, which shows up as an extra term in canonical Ward
identities describing formally the (softly broken) scale invariance of a quantum field theory model.
The formal theory of broken scale  invariance has been a classic theme in field theory since the
early 1970's, so we may perhaps start the subsequent discussion by writing down immediately a
typical Ward identity, referring for a general background to the standard literature
(see e.g. \cite{CJ,CCJ,C,IZ}). For the sake of technical simplicity, we will stay within the framework of spinor QED, at the
level of one-loop Feynman graphs.

  Let us consider a simple Ward identity involving the correlation function of the trace of energy-
momentum tensor and two electromagnetic currents (cf. \cite{ChE}), which reads, in its naive canonical
form
\begin{equation}
\left(2-p\cdot\frac{\partial}{\partial p} \right) \Pi_{\mu\nu}(p)
= \Delta_{\mu\nu}(p)
\end{equation}
Here the $\Pi_{\mu\nu}$  stands for the vacuum polarization tensor
\begin{equation}
\Pi_{\mu\nu}(p) =
i \int d^4x e^{ipx} \langle 0| T(J_\mu(x) J_\nu(0)) |0\rangle
\end{equation}
and the $\Delta_{\mu\nu}$ represents the three point vertex function
\begin{equation}
\Delta_{\mu\nu}(p) =
 \int d^4x d^4y e^{ipy} \langle 0| T(\theta_\alpha^\alpha(x) J_\mu(y)
J_\nu(0)) |0\rangle
\end{equation}
where $\theta_\alpha^\alpha$ means the trace of the ("improved")
energy-momentum tensor, equal to the divergence
of the dilatation current \cite{CCJ}. In spinor QED one has, in the lowest order
\begin{equation}
\theta_\alpha^\alpha(x) = m \bar\psi(x) \psi(x)
\end{equation}
The relation (9) is a "zero-energy theorem", since the divergence of dilatation current (and hence
the mass insertion in (11)) is taken at zero external momentum. It can be derived formally by
means of the naive canonical manipulations with the correlation function involving the dilatation
current and reflects thus the scale invariance of QED, softly broken by the fermion mass term (12).
Let us recall that the expression on the left-hand side of (9) comes from the commutators of
currents with the dilatation charge, while the right-hand side corresponds to the classical
symmetry-breaking term.
It is also useful to remember that the mass insertion at zero momentum can formally be produced
by means of  a differentiation with respect to mass, so that
\begin{equation}
m\frac{\partial}{\partial m} \Pi_{\mu\nu}(p;m)
= \Delta_{\mu\nu}(p;m)
\end{equation}
The current conservation implies transversality of the
$\Pi_{\mu\nu}$ and $\Delta_{\mu\nu}$, so one may define the
corresponding formfactors by
\begin{eqnarray}
\Pi_{\mu\nu}(p) = \Pi(p^2) (p_\mu p_\nu - p^2 g_{\mu\nu})
\nonumber\\
\Delta_{\mu\nu}(p) = \Delta(p^2) (p_\mu p_\nu - p^2 g_{\mu\nu})
\end{eqnarray}
(notice that the $\Pi$ and $\Delta$ are dimensionless). The naive
Ward identity (9) may then be recast as
\begin{equation}
-2p^2\frac{\partial}{\partial p^2} \Pi(p^2;m^2)
= \Delta(p^2;m^2)
\end{equation}

  Of course, in arriving at (15) one ignores any problems connected with ultraviolet divergences of
the underlying Feynman diagrams. Although (15) represents a relation between two finite
quantities, these should be defined properly, e.g. by means of  an appropriate gauge invariant
ultraviolet regularization. When this is done, the naive identity picks up an additional term - the
trace anomaly \cite{CJ,ChE}. At one-loop level one gets
\begin{equation}
-2p^2\frac{\partial}{\partial p^2} \Pi(p^2;m^2)
= \Delta(p^2;m^2) + \frac{1}{6\pi^2} e^2
\end{equation}
with $e$ being the electromagnetic coupling constant. The identity (16)
exemplifies the widely known connection of the trace anomaly with
the charge renormalization beta function \cite{ACD} (let us recall
that such a connection becomes rather transparent when considering
the limit $p^2 \to \infty$ in (16) and taking into account the dominance
of logarithmic asymptotics of the $\Pi(p^2)$ over the
$\Delta(p^2)$ ).

  We will now show that the anomalous term in (16) can be recovered
through dispersion relations, and expressed in terms of the imaginary
part of the formfactor $\Delta(p^2)$ (obviating thus the problems of
ultraviolet  nature), in close analogy with the case of axial
anomaly described in the preceding section. To this end, it is natural
to define the derivative of the $\Pi(p^2)$ by means of a differentiated
dispersion relation
\begin{equation}
\frac{\partial}{\partial p^2} \Pi(p^2)
= \frac{1}{\pi} \int_{4m^2}^{\infty}
\frac{\partial}{\partial p^2} \frac{\im\Pi(t)}{t-p^2} dt
\end{equation}
It is easy to see that (17) also follows readily from the usual (once subtracted)
dispersion relation for the $\Pi(p^2)$
\begin{equation}
\frac{1}{p^2} \Pi(p^2)
= \frac{1}{\pi} \int_{4m^2}^{\infty}
\frac{\im\Pi(t)}{t-p^2}\frac{dt}{t}
\end{equation}
Similarly, the $\Delta(p^2)$ can be defined by an unsubtracted dispersion relation
\begin{equation}
\Delta(p^2)
= \frac{1}{\pi} \int_{4m^2}^{\infty}
\frac{\im\Delta(t)}{t-p^2} dt
\end{equation}
In writing (17) and (19) we have simply assumed the necessary convergence properties of the
dispersion integrals, i.e. a proper asymptotic behaviour of the imaginary parts;  an explicit
illustration at the one-loop level is given below. The imaginary parts are well-defined finite
quantities and should therefore satisfy the canonical (non-anomalous) Ward identity
\begin{equation}
-2t\frac{\partial}{\partial t} \im\Pi(t;m^2)
= \im\Delta(t;m^2)
\end{equation}
To have some explicit one-loop expressions at hand, one may remember the familiar formula
\begin{equation}
\im\Pi(t;m^2) =
\frac{e^2}{12\pi}\left(1+\frac{2m^2}{t}\right)
\sqrt{1-\frac{4m^2}{t}}
\end{equation}
for $t>4m^2$ (see e.g. \cite{IZ} ). Using (20), one immediately gets also
\begin{equation}
\im\Delta(t;m^2) =
-\frac{2e^2}{\pi} \frac{m^4}{t^2}
\frac{1}{\sqrt{1-\frac{4m^2}{t}}}
\end{equation}
(the last result can of course be checked independently by a Feynman graph calculation). Let us
now calculate the quantity on the right-hand side of (16). Using the
definition (17) and integrating by parts, one obtains first
\begin{eqnarray}
-2p^2\frac{\partial}{\partial p^2} \Pi(p^2)
&=& -2p^2 \frac{1}{\pi} \int_{4m^2}^{\infty}
\frac{1}{(t-p^2)^2}\im\Pi(t) dt
\nonumber\\
&=& -2p^2 \frac{1}{\pi} \int_{4m^2}^{\infty}
\frac{1}{t-p^2}\frac{\partial}{\partial t}\im\Pi(t) dt
\end{eqnarray}
In arriving at the last expression we have dropped the corresponding
surface term; this is justified in view of the boundary values of
the $\im \Pi(t)$ (cf. (21)). In (23) one may now employ the identity
(20) to get, after a simple manipulation
\begin{eqnarray}
-2p^2\frac{\partial}{\partial p^2} \Pi(p^2)
&=& p^2 \frac{1}{\pi} \int_{4m^2}^{\infty}
\frac{\im\Delta(t)}{t(t-p^2)} dt
\nonumber\\
&=& \frac{1}{\pi} \int_{4m^2}^{\infty}
\frac{\im\Delta(t)}{t-p^2} dt
- \frac{1}{\pi} \int_{4m^2}^{\infty}
\frac{\im\Delta(t)}{t} dt
\end{eqnarray}
Taking into account the definition (19) one may thus write finally
\begin{equation}
-2p^2\frac{\partial}{\partial p^2} \Pi(p^2)
= \Delta(p^2)
- \frac{1}{\pi} \int_{4m^2}^{\infty}
\frac{\im\Delta(t)}{t} dt
\end{equation}
Having reproduced the form of the anomalous Ward identity (16), one should check that the
correct value of the trace anomaly is indeed recovered in (25). Using the expression (22) and
performing the integral one obtains
\begin{equation}
\frac{1}{\pi} \int_{4m^2}^{\infty}
\frac{\im\Delta(t)}{t} dt
= -\frac{1}{6\pi} e^2
\end{equation}
so that the anticipated result (16) is confirmed. From (22) and (26) it is also clear that
\begin{equation}
\lim_{m\to 0} \frac{1}{t} \im \Delta(t;m^2) = -\frac{e^2}{6\pi}
\delta(t)
\end{equation}

  Thus, within the dispersion relation approach the trace anomaly is
tantamount to the sum rule (26) for the imaginary part of the classical
symmetry-breaking term $\Delta(t;m^2)$, and in the massless limit it
is manifested through the delta-function singularity of the relevant
imaginary part at $t=0$. A remarkable feature of such an approach is
that one obtains a quantity intimately related to the
charge-renormalization beta function (which is inherently of ultraviolet nature) without ever
mentioning an ultraviolet regularization. There is obviously a striking similarity between the results
found here and those described in the preceding section for the well-known axial anomaly. In
particular, it is worth emphasizing that in both cases the anomaly can actually be expressed in the
same way, namely through the imaginary part of the classical symmetry-breaking term : Indeed, in
the case of axial anomaly (cf. (5)) one has
$\im F_1(t) = 2m \im G(t)/t$  (see (4)), and the term $2mG(t)$ is clearly
a natural counterpart of the $\Delta(t)$ .

\section{Further remarks on the imaginary parts}

  Let us now add some further observations concerning the imaginary
parts of the considered correlation functions in the context of the
trace anomaly problem.

  First, it is easy to arrive at an alternative representation of
the trace anomaly in terms of $\im \Pi(p^2)$. Denoting the anomalous
term embodied in (25) by ${\cal A}$, i.e.
\begin{equation}
{\cal A}=-\frac{1}{\pi} \int_{4m^2}^{\infty}
\im\Delta(t;m^2)\frac{dt}{t}
\end{equation}
then employing the identity (20) one gets immediately
\begin{equation}
{\cal A}=\frac{2}{\pi} \int_{4m^2}^{\infty}
\frac{\partial}{\partial t} \im\Pi(t) dt
\end{equation}
and therefore
\begin{equation}
{\cal A}= \lim_{p^2 \to \infty} \frac{2}{\pi}
\im\Pi(p^2;m^2)
\end{equation}
Note that a relation of this type was in fact considered earlier (cf. \cite{ChE}, where it was deduced in a
somewhat indirect way). Within our dispersive approach, the result (30) emerges as a
straightforward consequence of the basic integral representation (28).

  Another remarkable technical feature of the anomaly (28) is that it may be recast as
\begin{equation}
{\cal A}= -\Delta(0;m^2)
\end{equation}
The relation (31) has also been discussed in the early days of the
anomaly theory (see \cite{ChE} ) by  means of different (short-distance)
methods. Within our approach it becomes obvious immediately,
by comparing (28) with the definition (19). In fact, (31) is easily
understood if one takes into account that the $\Pi(p^2;m^2)$ has
no singularity at $p^2 = 0$ for $m \not= 0$ (note, however, that in the
massless limit the $\Pi(p^2;0)$ does have a logarithmic infrared
singularity, which reproduces precisely the trace anomaly in (16)).
In explicit terms, (31) means that the trace anomaly is determined by
the value of the three point function formfactor $\Delta$  with all
external momenta set to zero. In our calculation, the external momentum
attached to the mass insertion is set to zero from the start and
then the limit $p^2 \to 0$ can be taken. As an additional check of
our approach, one would like to see whether a limiting procedure
taken in reverse order would produce
the same result. To clarify this point, let us consider the relevant
three point function (denoted here as $\tilde\Delta_{\mu\nu}$
to distinguish it from the previous kinematical configuration)
with light-like momenta $k, p$ attached to the currents;
the corresponding formfactor is then
a function of the kinematical variable $q^2$, with $q$ being
the four-momentum attached to the mass-insertion vertex
(i.e. $q^2 = (k + p)^2$ ). Again, one may write an unsubtracted
dispersion relation for the $\tilde\Delta(q^2;m^2)$, i.e.
\begin{equation}
\tilde\Delta(q^2;m^2)
= \frac{1}{\pi} \int_{4m^2}^{\infty}
\frac{\im\tilde\Delta(s)}{s-q^2} ds
\end{equation}
A straightforward calculation now yields
\begin{equation}
\im \tilde\Delta(s;m^2) = -\frac{e^2}{8\pi}\frac{4m^2}{s}
\left(1-\frac{4m^2}{s} \right)
\ln\frac{1+\sqrt{1-\frac{4m^2}{s}}}{1-\sqrt{1-\frac{4m^2}{s}}}
\end{equation}
 The quantity
\begin{equation}
\tilde\Delta(0)= \frac{1}{\pi} \int_{4m^2}^{\infty}
\im\Delta(s;m^2)\frac{ds}{s}
\end{equation}
is then expected to be equal to the $\Delta(0)$ calculated before (we
have suppressed the dependence on $m^2$ as it is obviously trivial for
zero external virtualities). The relevant integral occurring in (34)
may be easily evaluated as
\begin{equation}
\int_1^\infty
\frac{1}{y} \left(1-\frac{1}{y} \right)
\ln\frac{1+\sqrt{1-\frac{1}{y}}}{1-\sqrt{1-\frac{1}{y}}}
\frac{dy}{y}
= \frac{4}{3}
\end{equation}
which then obviously reproduces  the anticipated result
\begin{equation}
\tilde\Delta(0)=\Delta(0)
\end{equation}
(cf.(26)) and an equivalence of the $t$-channel and $s$-channel
dispersive calculations of the trace anomaly is thus verified.

  The last remark concerns a possible generalization of the preceding
discussion to other field theory models. The previous results can be
extended to the scalar electrodynamics without any major changes. In a
previous paper \cite{HS} we have discussed briefly the problem of
a dispersive  derivation of the trace anomaly due to the $W$ boson
loops within the standard model of electroweak interactions, employing
the $s$-channel dispersion relations (in the sense indicated above)
in connection with the Higgs boson decay into two photons. The corresponding
argument becomes somewhat obscured by technical complications due to
the delicate nature of massive vector bosons in spontaneously broken
gauge theories and a corresponding $t$-channel calculation (which would
be analogous to the main line of the present paper) also requires special
care. In general, the problem of dispersive derivation of the trace
anomaly in non-Abelian gauge theory models would deserve a separate
treatment.

\section{Summary}

  Let us now summarize briefly the main results obtained in this paper. Invoking the paradigm of
the well-known axial anomaly, we have discussed a simple derivation of the trace anomaly in
spinor QED by means of dispersion relations. The trace anomaly, defined as an extra term in a
canonical Ward identity (zero-energy theorem) for the broken scale invariance, has been expressed
as a convergent sum rule for the imaginary part of the classical symmetry-breaking term (i.e. as an
integral along the cut associated with the non-zero imaginary part of a relevant formfactor). In the
massless limit, one gets a delta-function singularity at zero external momentum squared, which
reveals an alternative "infrared face" of the trace anomaly. We have checked explicitly that the
same result for the anomaly is recovered when employing dispersion relations in different
kinematical variables ($t$- and $s$-channel resp.). The results obtained here exhibit a striking similarity
with the case of the axial anomaly, for which the dispersive infrared approach was pioneered by
Dolgov and Zakharov many years ago \cite{DZ}.

  In general, the infrared (low-energy) aspects of quantum anomalies are not discussed frequently
in the literature, and for various reasons the ultraviolet (short-distance) nature of these phenomena
is usually emphasized. For the trace anomaly, its remarkable connection with the coupling-constant
renormalization beta function is certainly the best known (ultraviolet) aspect. Nevertheless, it may
be useful to know its dispersive derivation and the associated infrared face as well, in analogy with
the archetypal example of the axial anomaly.

  Finally, let us remark that some of the points emphasized in the
present paper were also mentioned briefly, in a slightly different
context, in the recent paper \cite{D} by Deser. Another
independent investigation in this direction (for a scalar field
model) has recently been communicated to us by O. Teryaev
\cite{T}.

\smallskip
{\bf Acknowledgements :} We are grateful to J. Novotn\'y for useful discussions. One of us (J.H.) thanks
R. Bertlmann for an encouraging discussion during the Triangle seminar at Vienna University. The
work has been supported in part by research grants GAUK-166/95 and GACR-1460/95.

\end{document}